\documentclass[conference]{IEEEtran}

\usepackage{amsmath}
\usepackage{graphicx}
\usepackage{subfig}
\usepackage{caption}
\usepackage{verbatim}

\graphicspath{{/}}

\begin{document}

\title{Grouping Entities in a Fleet by Community Detection in Network of Regression Models}

\author{
\IEEEauthorblockN{Pankaj Pansari}
\IEEEauthorblockA{General Electric\\
Bangalore, India\\
pankaj.pansari@ge.com}
\and
\IEEEauthorblockN{C.Rajagopalan}
\IEEEauthorblockA{General Electric\\
Bangalore, India\\
rajagopalan.c@ge.com}
\and
\IEEEauthorblockN{Ramasubramanian Sundararajan \textsuperscript{\footnotemark1}}
\IEEEauthorblockA{Sabre Airlines Solutions\\
Bangalore, India\\
ramasubramanian.sundararajan@sabre.com}}

\maketitle

\begin{abstract}

This paper deals with grouping of entities in a fleet based on their behavior. The behavior of each entity is characterized by its historical dataset, which comprises a dependent variable, typically a performance measure, and multiple independent variables, typically operating conditions. A regression model built using this dataset is used as a proxy for the behavior of an entity. The validation error of the model of one unit with respect to the dataset of another unit is used as a measure of the difference in behavior between two units. Grouping entities based on their behavior is posed as a graph clustering problem with nodes representing regression models and edge weights given by the  validation errors. Specifically, we find communities in this graph, having dense edge connections within and sparse connections outside. A way to assess the goodness of grouping and finding the optimum number of divisions is proposed. The algorithm and measures proposed are illustrated with application to synthetic data.

\end{abstract}

\section{Introduction}

Monitoring of infrastructure such as aircrafts, turbines and vehicles is becoming increasingly important. The sensors on these entities generate a wealth of data, which is analysed to predict failures in them. Typically the analysis of historical data has been on an entity-by-entity basis \cite{nielsen06}. However, given a fleet of units, it is possible to pose questions about the collective behavior of the fleet and answer them using the tools of network analysis. One such question pertains to finding units in the fleet which behave similarly.

Specifically, our problem is to cluster sets of measurements \(Y\) measured as a function of dependent variable(s) $x$. Data, in the form of discrete observations, is given for $N$ different entities and for each entity, $Y$ is assumed to be a smooth function of $x$. The length of dataset need not be same for all entities. In cases where $x$ is time measurement, the data is referred to as \emph{longitudinal} data or \emph{repeated measures} data. In a general context, where $x$ need not be time, the term \emph{functional} or \emph{trajectory} data is used. Given trajectory data, we are interested in determining if the data can be naturally clustered into groups. 

\footnotetext[1]{This work was done when the author was at General Electric Global Research in 2014.}

Major functional data clustering approaches can be classified in one of three groups: two-stage methods, non-parametric clustering and model-based clustering \cite{jacques13}. These algorithms aim to cluster \emph{curves}, where it is implied that the dimensionality of data is very large but no relation between $Y$ and $x$ is utilized. Hence, these algorithms do not characterize the entities by their behavior or by the way in which $Y$ depends on $x$.

Catez \emph{et al.}\cite{cadez00} have clustered trajectories by explicitly using the depedency of \textbf{Y} on $x$ using the expectation-maximization (EM) algorithm to find the grouping. It is based on probabilistic modeling of a set of trajectories generated from a finite mixture model consisting of regression model components. The algorithm proceeds by making an assumption about the functional form of the component models. Our method, on the other hand, does not involve such an assumption, which may be difficult make a priori in several cases.

Given the historical data of a unit, our method first computes a regression model for each unit. This model stands as a proxy for the behavior of the unit. The fleet is represented by an ensemble of such models, one for each unit. We propose a suitable dissimilarity measure between two units and form a graph of the fleet. Each vertex of the graph represents the model of a unit and the edge weight between two vertices is given by the value of dissimilarity measure between two models. The trajectory clustering problem is thus translated to a graph clustering problem. We also propose a measure to characterize the goodness of grouping of similarly-behaving units and hence find the best division.

Section II gives a formal description of the problem and precise meaning of terms used above - behavior, adequate modeling accuracy and similarity of behavior. It also gives the method used to find optimal grouping of the units. Section III demonstrates the application of our approach for performance modeling of a fleet of steam turbines. Section IV discusses potential alternative approaches and directions for future work.

\begin{figure*}[ht!]
        \centering
        \subfloat[Original Graph]{
                \includegraphics[width=0.4\textwidth]{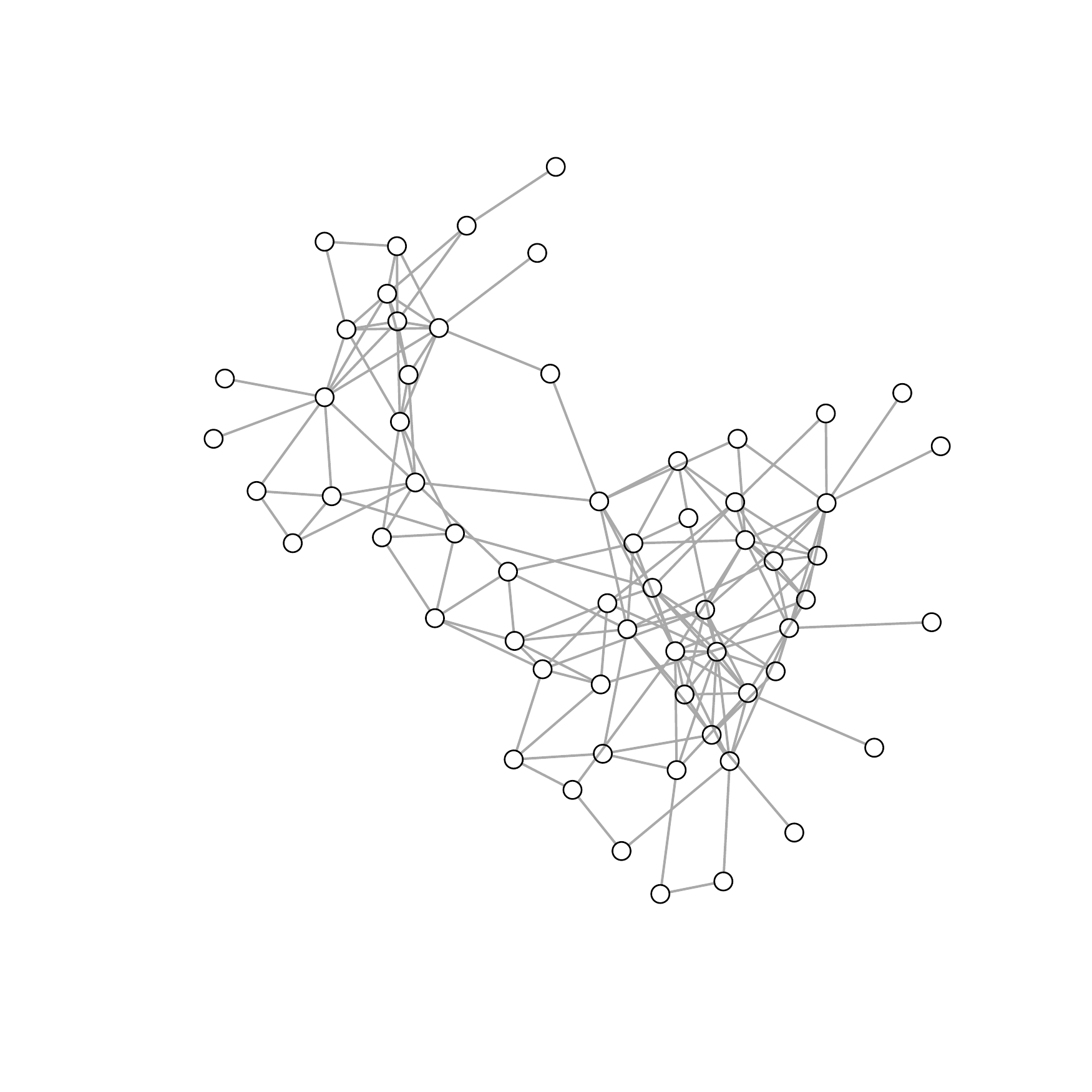}
        		
        }
        \quad
        \subfloat[Clusters]{
                \includegraphics[width=0.4\textwidth]{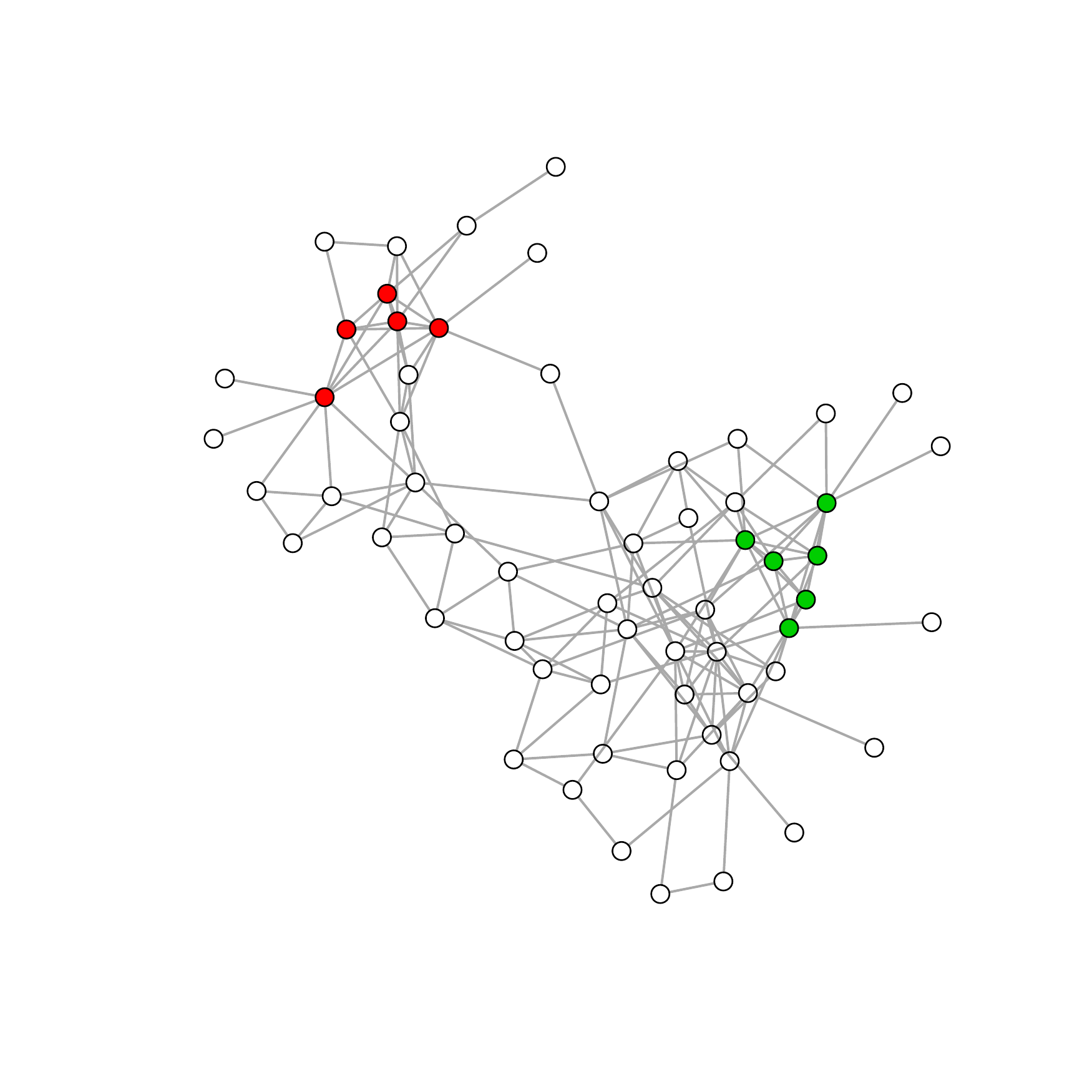}
                
        }
        \quad
        \subfloat[Connected components]{
                \includegraphics[width=0.4\textwidth]{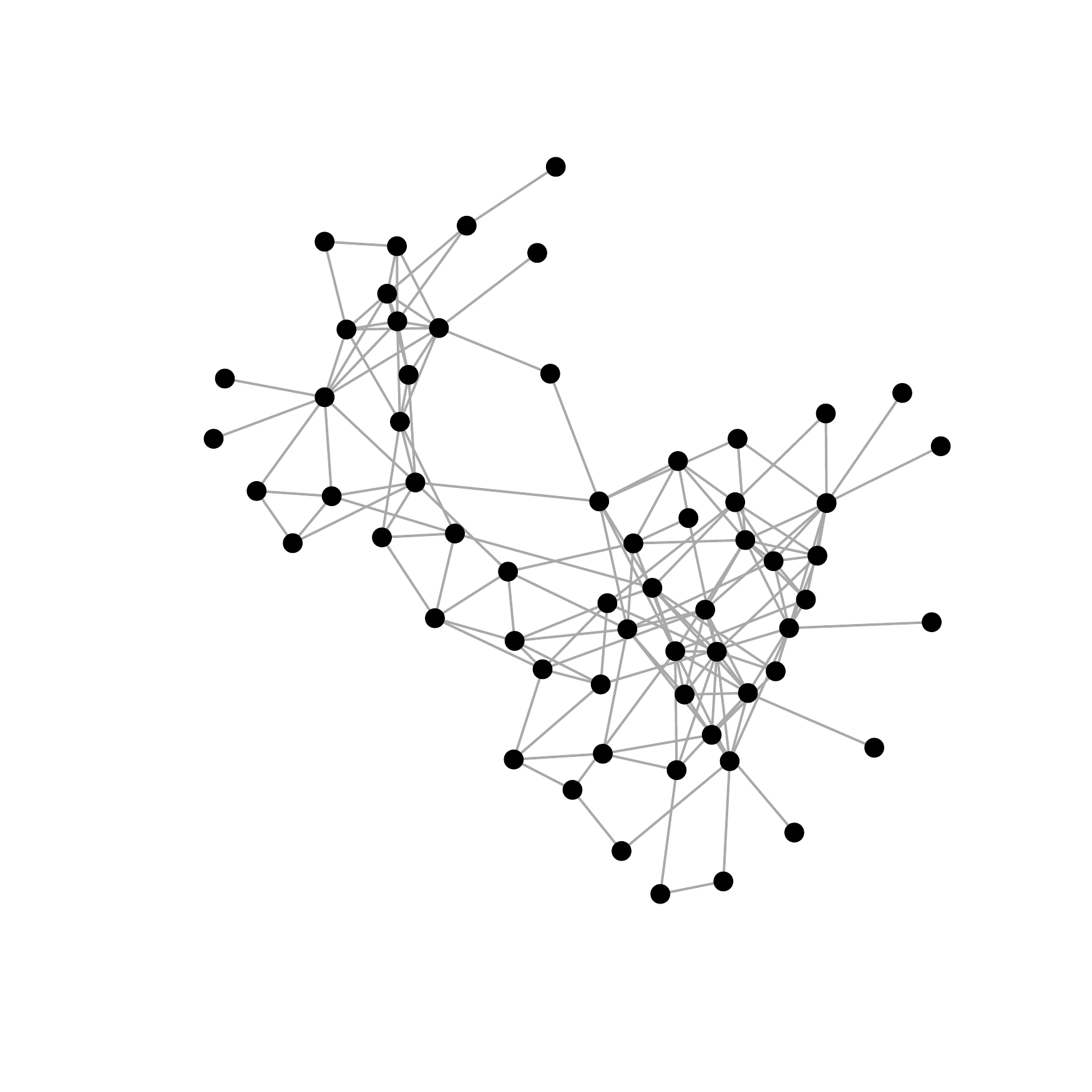} 
        		
        }
        \quad
        \subfloat[Communities]{
                \includegraphics[width=0.4\textwidth]{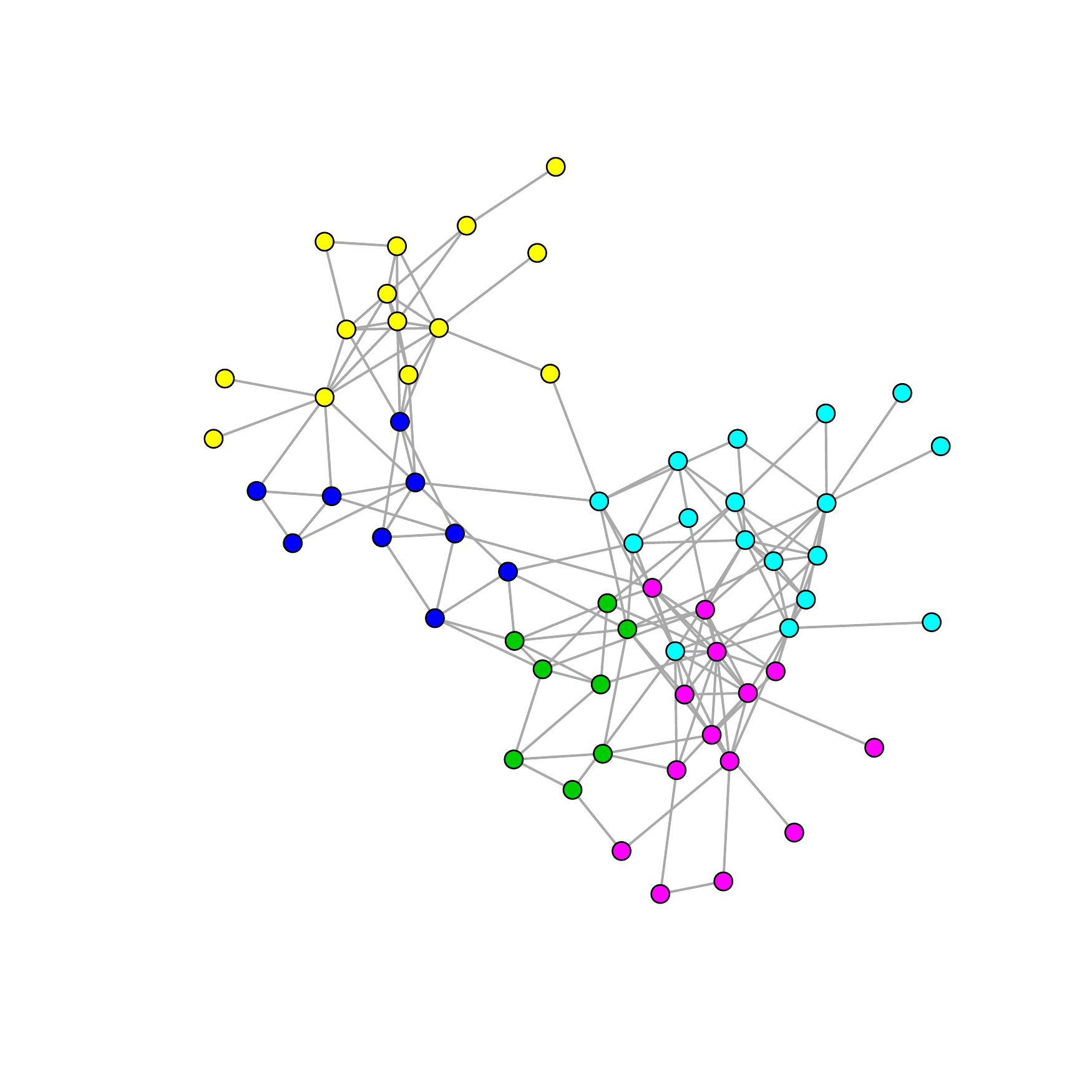}
        		
        }
        
        \caption{Different kinds of groups in a network. (a) shows the undirected social network of frequent associations among 62 dolphins in a community living off Doubtful Sound, New Zealand \cite{lusseau03}. (b) and (c) depict cliques and connected component respectively in the network. A clique is too strict  a notion of a group and a connected component too loose with neither providing much meaningful insight. (d) shows communities which are densely connected within and sparsely connected without. For our purpose, a community is a suitable representation of the notion of a group.}
        \label{fig:groups}
\end{figure*}

\section{Method}

\subsection{Model Estimation of Entities}

Given the hisorical dataset of the fleet in the form

\begin{equation}
\left\{\{X^{i}(t), Y^{i}(t)\}_{t=1}^{T_{i}}\right\}_{i=1}^{N}
\end{equation}

where $i$ numbers the unit in the fleet, and the integer $t$ denotes the occasion (time) when the data from unit $i$ is recorded. The dataset of each unit has length $T_{i}$ occasions (time samples). The independent variables $X(t)$ $\in$ $\textbf{R}^{p}$ are the input measurements - measurements of the operating conditions of unit $i$ at time $t$. The dependent variable $Y(t)$ $\in$ $\textbf{R}$ is the output measurement - typically, a performance metric of unit $i$ at time $t$ \cite{chu11}.

The historical data of unit $i$ is denoted as $D_{i}$. The \emph{behavior} of unit $i$ implies the relationship between $Y$ and $X = (X_{1}, X_{2}, ......, X_{p})$ which can be written in the general form

\begin{equation}
Y = f(X) + \epsilon
\end{equation}

Here $f$ is some fixed but unknown function of $X_{1}, ....., X_{p}$ and $\epsilon$ is  a random error term which is independent of $X$ and has zero mean. Since $f$ is unknown, we compute an estimate $\hat{f}$ of $f$ from historical data. With $\hat{Y} = \hat{f}(X)$, we have

\begin{align}
E(Y - \hat{Y})^{2} &= E[f(X) + \epsilon - \hat{f}(X)]^{2} \notag \\
 &= \underbrace{E[f(X) - \hat{f}(X)]^{2}}_{\text{Reducible}} + \underbrace{Var(\epsilon)}_\text{Irreducible}
\end{align}

where $E(Y - \hat{Y})^2$ represents the average, or \emph{expected value}, of the squared difference between the predicted and actual value of $Y$, and Var($\epsilon$) represents the variance associated with the error term $\epsilon$ \cite{james13}. 

Given $D_{i}$, $f_{i}$ can be estimated by either parametric or non-parametric statistical methods. If a parametric method is used, the functional form of $f_{i}$ may be assumed linear or non-linear, depending on the application and the dataset $D_{i}$. For a fleet of $N$ units, we have an ensemble of $N$ models - $f_{1}$, $f_{2}$...$f_{N}$. To assess the accuracy of the models, residual standard error (RSE) can be used and is given by

\begin{equation}
RSE = \sqrt{\frac{1}{T_{i}}\sum_{t=1}^{T_{i}} (Y(t) - \hat{Y}(t))^{2}}
\end{equation}

\subsection{Constructing Fleet Network of Regression Models}

The RSE is a measure of the lack of fit of the model to the data. The same measure can be used to represent the dissimilarity between two models. Let $r_{ij}$ stand for the RSE when model $f_{i}$ is tested on the dataset $D_{j}$. Given the estimated models $f_{1}$, $f_{2}$...$f_{N}$ and the datasets $D_{1}$, $D_{2}$...$D_{N}$, a $N \times N$ matrix $\textbf{R}$ of RSE errors can be constructed

\begin{equation}
\textbf{R} = \begin{pmatrix}
 r_{11} & r_{12} & \cdots & r_{1N} \\
 r_{21} & r_{22} & \cdots & r_{2N} \\
 \vdots  & \vdots  & \ddots & \vdots  \\
 r_{N1} & r_{N2} & \cdots & r_{NN}
\end{pmatrix}
\end{equation}

\textbf{R} represents a weighted multigraph. For simplicity of analysis, we would like to consider weighted, simple graphs. This can be accomplished by

\begin{align}
\textbf{R}' &= \frac{\textbf{R} + \textbf{R}^T}{2} \\ \notag \\
			&= \begin{pmatrix}
 				r'_{11} & r'_{12} & \cdots & r'_{1N} \\ \notag
 				r'_{21} & r'_{22} & \cdots & r'_{2N} \\ \notag
 				\vdots  & \vdots  & \ddots & \vdots  \\ \notag
 				r'_{N1} & r'_{N2} & \cdots & r'_{NN} \notag
				\end{pmatrix}
\end{align}

$\textbf{R}'$ is a symmetric matrix, that is, $r'_{ij}$ = $r'_{ji}$. Also, $r'_{ii}$ represents the training error of $f_{i}$. $\textbf{R}'$ can be considered as a dissimilarity matrix, where $r'_{ij}$ and $r'_{ji}$ can be considered as a measure of difference between units $i$ and $j$.\\

Consider a transformation of $\textbf{R}'$ to $\textbf{A}(\lambda)$ given by

\begin{equation}
\textbf{A}(\lambda) = \begin{pmatrix}
						0 & H(\lambda - r'_{12}) & \cdots & H(\lambda - r'_{1N}) \\
                        H(\lambda - r'_{21}) & 0 & \cdots & H(\lambda - r'_{2N})\\
                        \vdots  & \vdots  & \ddots & \vdots  \\
                      H(\lambda - r'_{N1}) & H(\lambda - r'_{N2}) & \cdots & H(\lambda - r'_{NN})  
					  \end{pmatrix}
\end{equation}

where $\lambda$ is a given scalar value and H(x) is the unit step function given by

\begin{equation}
H(x) = 
	\begin{cases}
	0 & \text{if } n < 0, \notag \\
	1 & \text{if } n \geq 0
    \end{cases}
\end{equation}

$\textbf{A}(\lambda)$ is a binary, symmetric matrix. It can be considered as an adjacency matrix of a graph. This graph has an edge between a pair of vertices $V_{i}$ and $V_{j}$ (representing units $i$ and $j$ respectively) only if 
$r'_{ij}$, or equivalently $r'_{ji}$, is less than the given threshold $\lambda$ with self-loops being removed.

\subsection{Grouping of Entities in Network}

The notion of groups in a graph may be represented by three concepts: (a) clique (b) community (c) connected component.

The kind of networks we are interested in possess community structure. This implies that the network divides naturally into groups of nodes with dense connections internally and sparser connections between groups. This same structural feature is an important property of social networks, such as those of dolphins illustrated in Fig. 1 \cite{newman03}.

Detecting cliques is analogous to complete-linkage clustering and finding connected components is akin to single-linkage clustering. For the networks being studied here, cliques are too strong a criterion and connected components too weak. In either case, we do not arrive at a meaningful grouping or obtain non-trivial knowledge. Community detection, being representative of average-linkage clustering, provides just the right balance between the two extremes. This fact is highlighted in Fig. 1.\\

\subsubsection{Detection of Communities}

Community structure detection algorithms try to find dense subgraphs in directed or undirected graphs, by optimizing a criterion particular to the algorithm. Several community detection algorithms have been developed, which are based on centrality measures, flow models, random walks, resistor networks, optimization and other approaches \cite{danon05}.

For our application, we have used the leading eigenvector method which has proven very effective \cite{newman06}. The heart of the method is the definition of a modularity matrix \textbf{B} 

\begin{equation}
\textbf{B} = \textbf{A}(\lambda) - \textbf{P}
\end{equation}

where $\textbf{A}(\lambda)$ is the adjacency matrix obtained in (7), and \textbf{P} contains the probability that certain edges are present according to the ‘configuration model’. In other words, a $\textbf{P}[i,j]$ element of \textbf{P} is the probability that there is an edge between vertices $i$ and $j$ in a random network in which the degrees of all vertices are the same as $\textbf{A}(\lambda)$. 

The leading eigenvector method works by calculating the eigenvector of the modularity matrix for the largest positive eigenvalue and then separating vertices into two community based on the sign of the corresponding element in the eigenvector. If all elements in the eigenvector are of the same sign that means that the network has no underlying comuunity structure. Further details can be found in \cite{newman06}.\\

\subsubsection{Finding Optimal Grouping}

Given a grouping by the preceding method, a measure to assess the accuracy of the grouping is required to ascertain the goodness of grouping. Such a measure, called average meta-validation accuracy, is described now.

For a given $\lambda$, let the community detection algorithm result in $k$ groups. We shall use the prefix \emph{meta} in reference to this grouping, thereby implying a higher level of abstraction and analysis. The datasets of all the units belonging to the $i$-th group are merged in a single meta-dataset $G_{i}$.  Hence, we have $k$ meta-datasets $G_{1}, G_{2}, ...., G_{k}$. Each $G_{i}$ is divided into a training set $G_{Ti}$ and a validation set $G_{Vi}$ in an suitable proportion.

Each $G_{i}$ can be considered to be generated by an underlying function $g(X)$, analogous to $f(X)$ in (2). Given the training set $G_{Ti}$ a meta-model $g_{i}$ can be estimated. The form of $g(X)$ can be parametric or non-parametric and can differ from that of $f(X)$. However, for the sake of comparison, the form of $g_{i}$ at the meta-level must be consistent for varying values of $\lambda$. Thus, we have an ensemble of meta-models $g_{1}, g_{2}, ...., g_{k}$.

For each model $g_{i}$, the validation RMSE $e_{i}$ is calculated using (4). Let $|G_{i}|$ represent the number of units in the group $G_{i}$. The average meta-validation accuracy, $\eta(\lambda)$ is defined as

\begin{equation}
\eta(\lambda) = 100 - \frac{\sum\limits_{i = 1}^{k}|G_{i}| \times e_{i}}{N}
\end{equation}

A plot of $\eta(\lambda)$ versus the number of communities detected (henceforth referred to as 'accuracy plot') tends to be a monotonically increasing curve. The point where the slope of the curve levels off (the 'elbow point') indicates the suitable number of communities present in the fleet.

\section{Results}

We analysed the historical data for a fleet of 65 GE steam turbines to model the power output of the units as a function of the operating conditions in the steady-state. The data consisted of sensor readings of temperature and pressure at various stages of the turbine and the power output. The data was sampled at one hour interval over a period of 5-10 years. Due to its proprietary nature, more description of the turbine data and corresponding results cannot be given here.

\begin{figure*}

        \begin{minipage}[h]{\textwidth}
        \centering
        \includegraphics[width=0.65\textwidth]{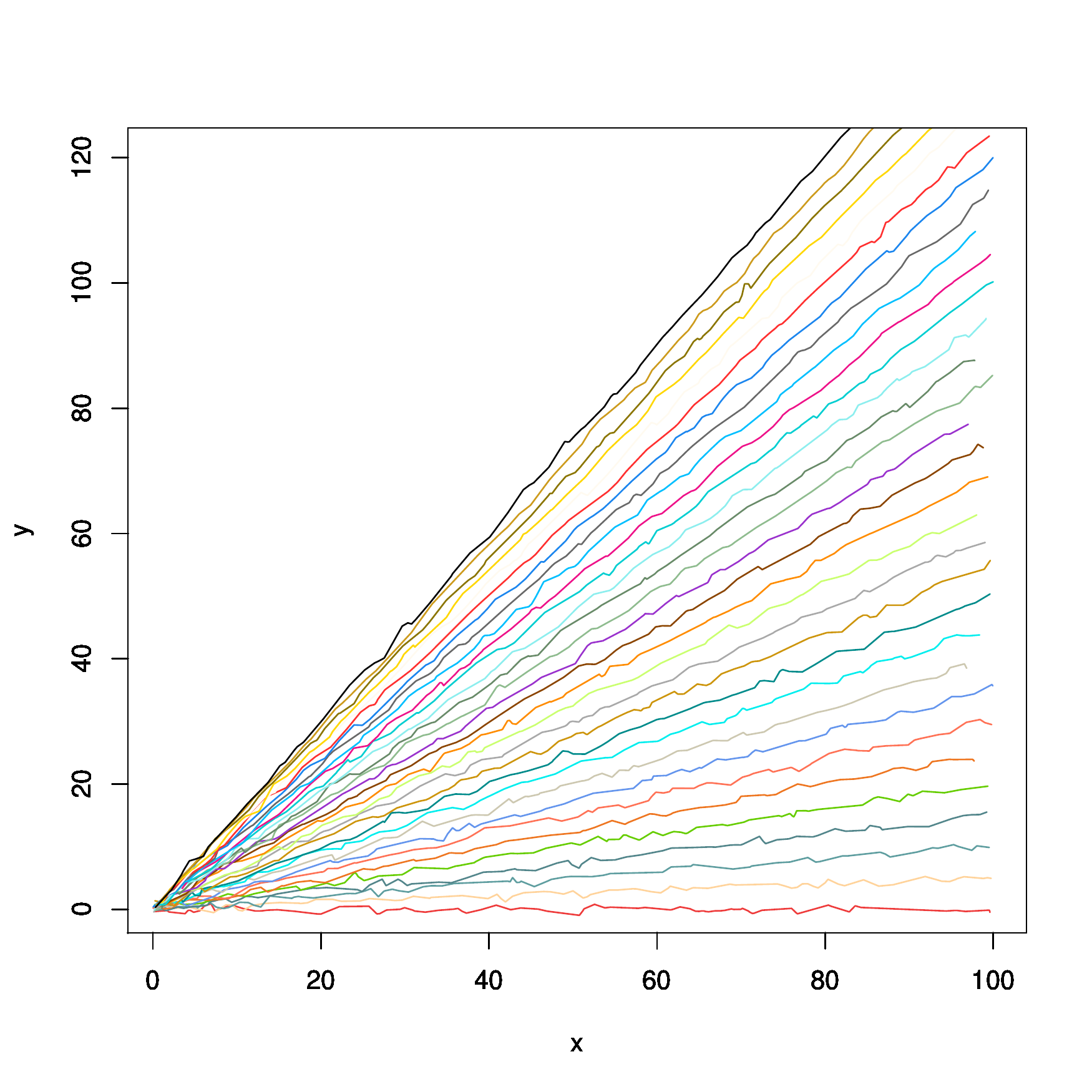}
        \end{minipage}
        \begin{minipage}[h]{\textwidth}
        \centering
        \includegraphics[width=0.65\textwidth]{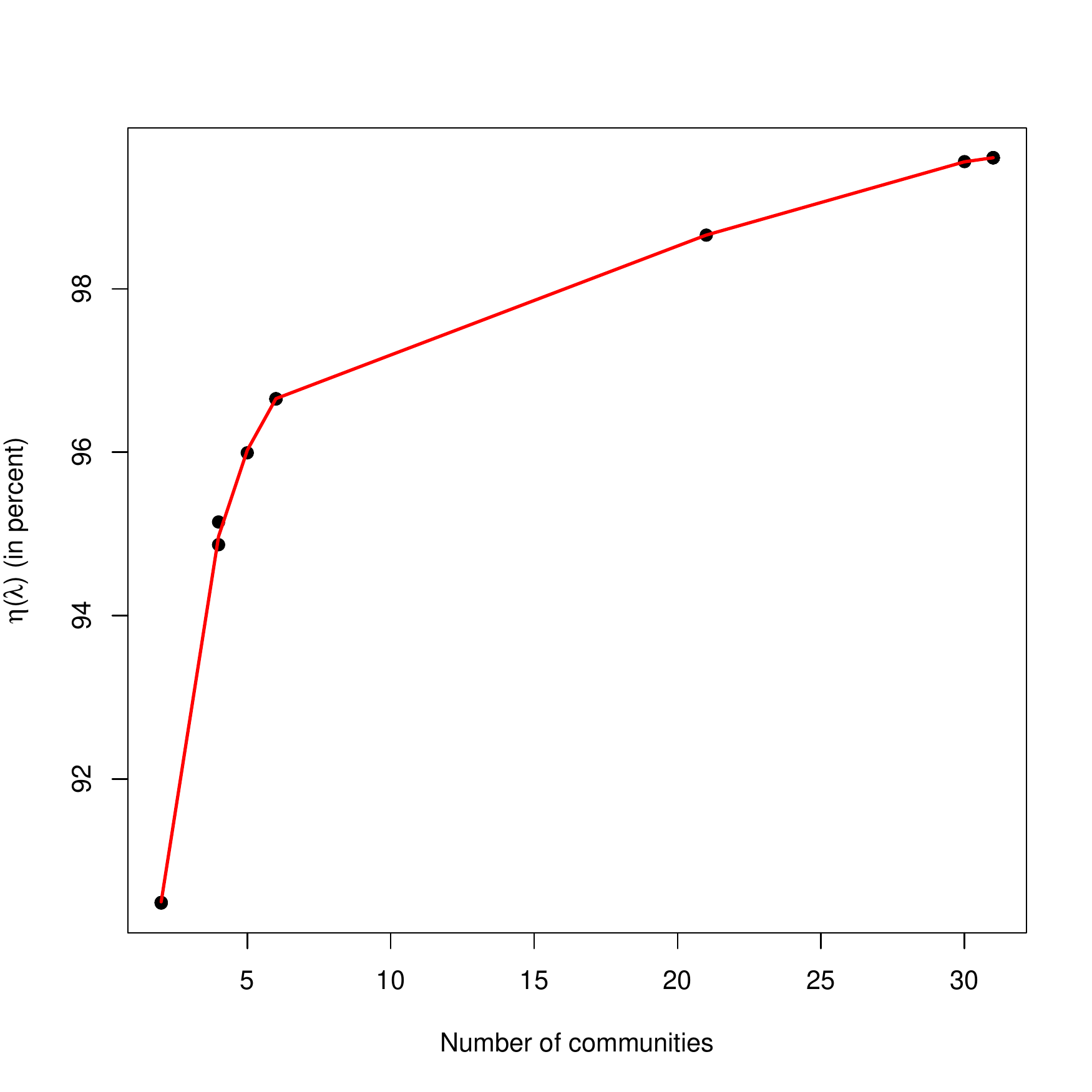}
        \end{minipage}
        \caption{Simulated data in which grouping does not exist. $\eta(\lambda)$ vs number of communities plot does not flatten off. The algorithm colors each curve differently. The population is best represented by collection of individual models.}
		\label{fig:no_groups}
        
\end{figure*}

\begin{figure*}        

		\begin{minipage}[b]{\textwidth}
        \centering
        \includegraphics[width=0.65\textwidth]{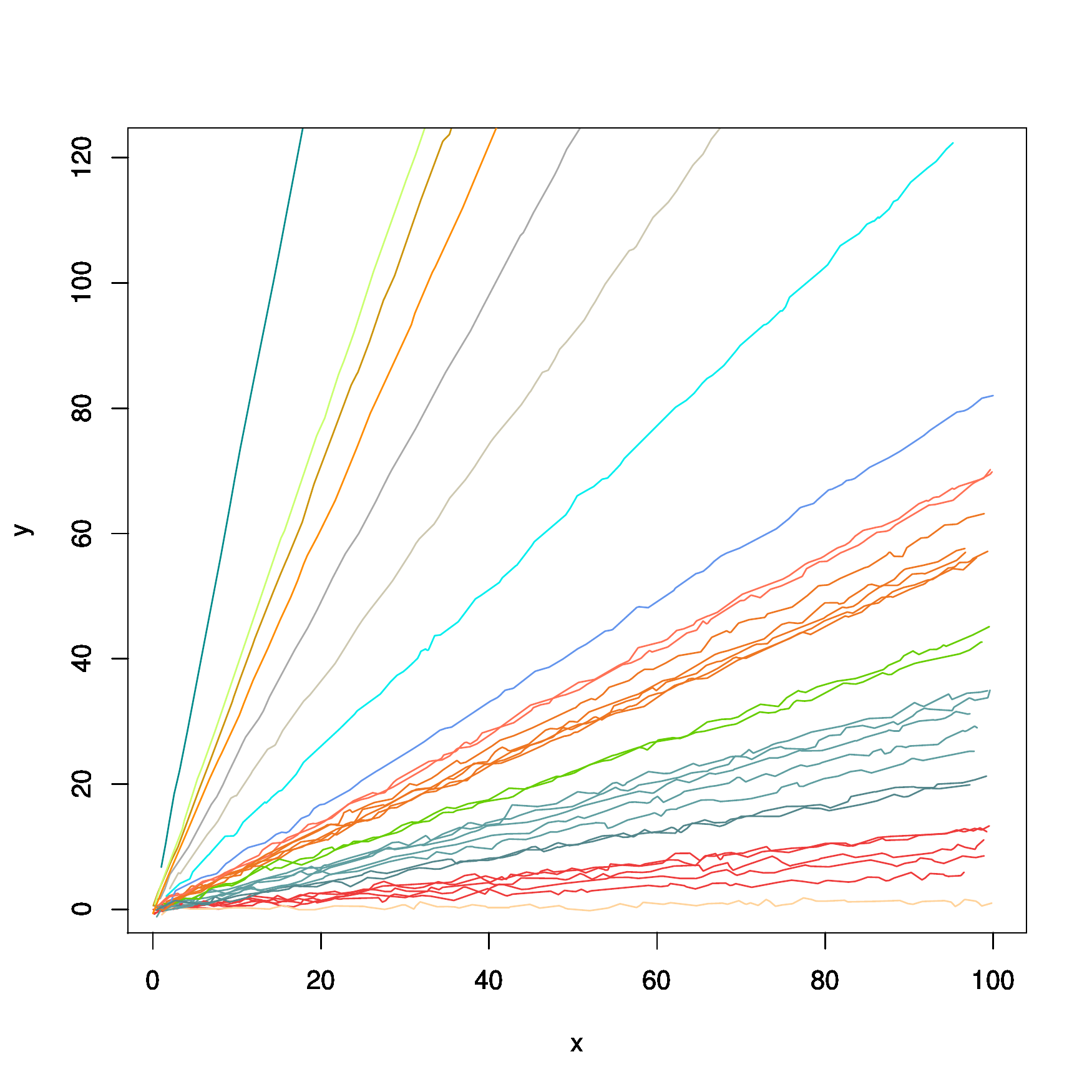}
        \end{minipage}
        \begin{minipage}[b]{\textwidth}
        \centering
        \includegraphics[width=0.65\textwidth]{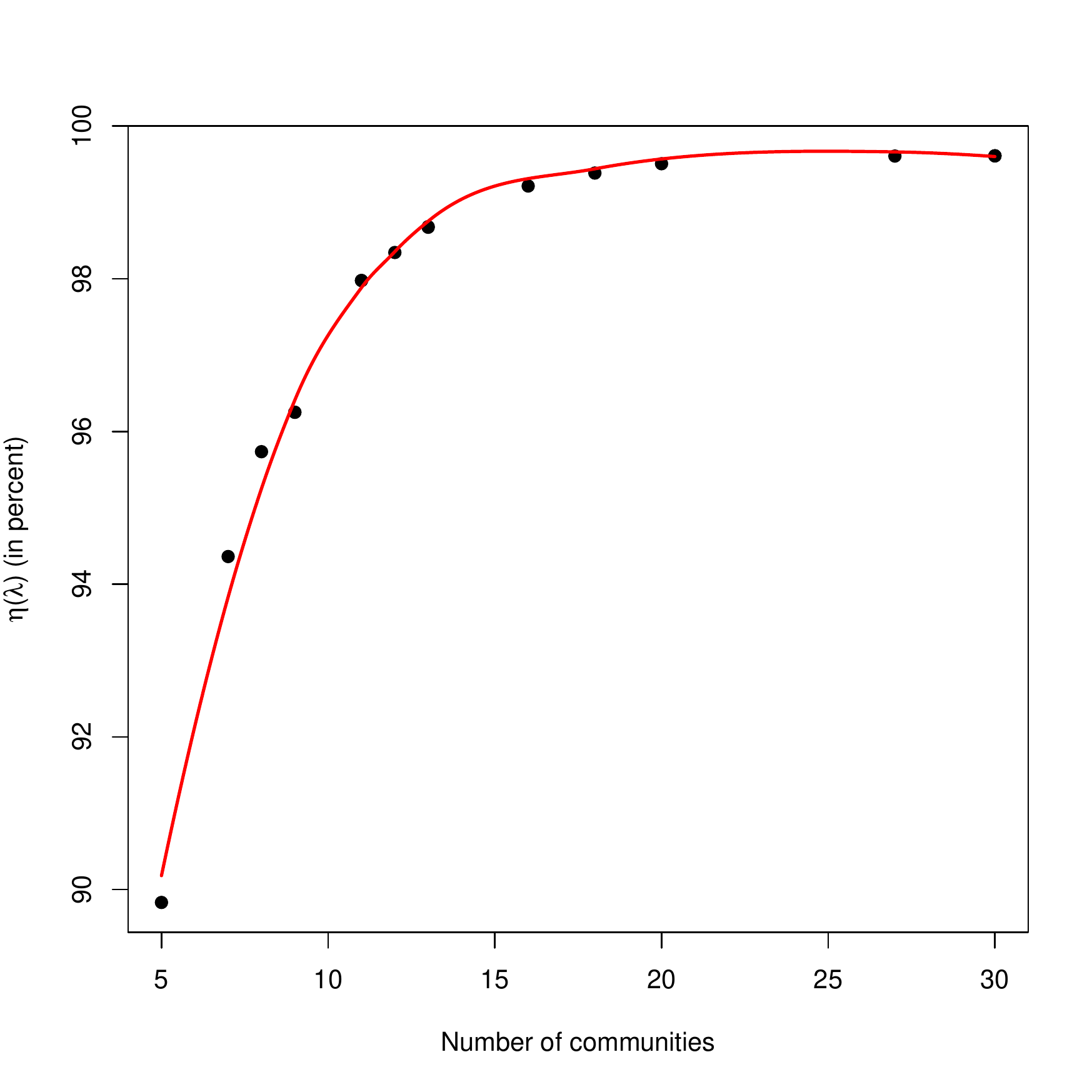}
        \end{minipage}
        \caption{Simulated data in which grouping is fuzzy.  $\eta(\lambda)$ vs number of communities plot flattens off, but not at a clear point. Curves with the same color are grouped together. The population can be well  represented by a collection of group models.}
		\label{fig:fuzzy_groups}

\end{figure*}

\begin{figure*}  

        \begin{minipage}[b]{\textwidth}
        \centering
        \includegraphics[width=0.65\textwidth]{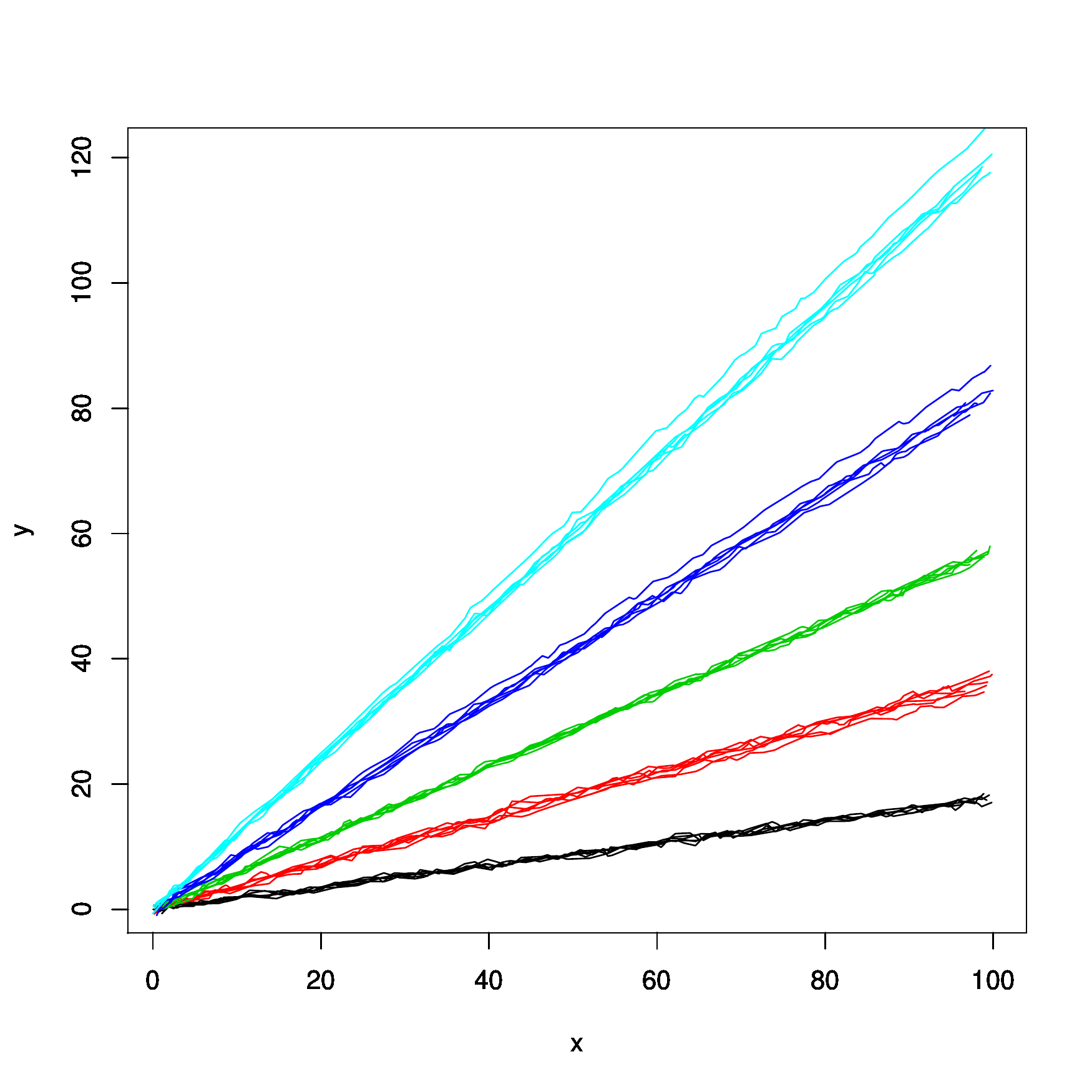}
        \end{minipage}
        \begin{minipage}[b]{\textwidth}
        \centering
        \includegraphics[width=0.65\textwidth]{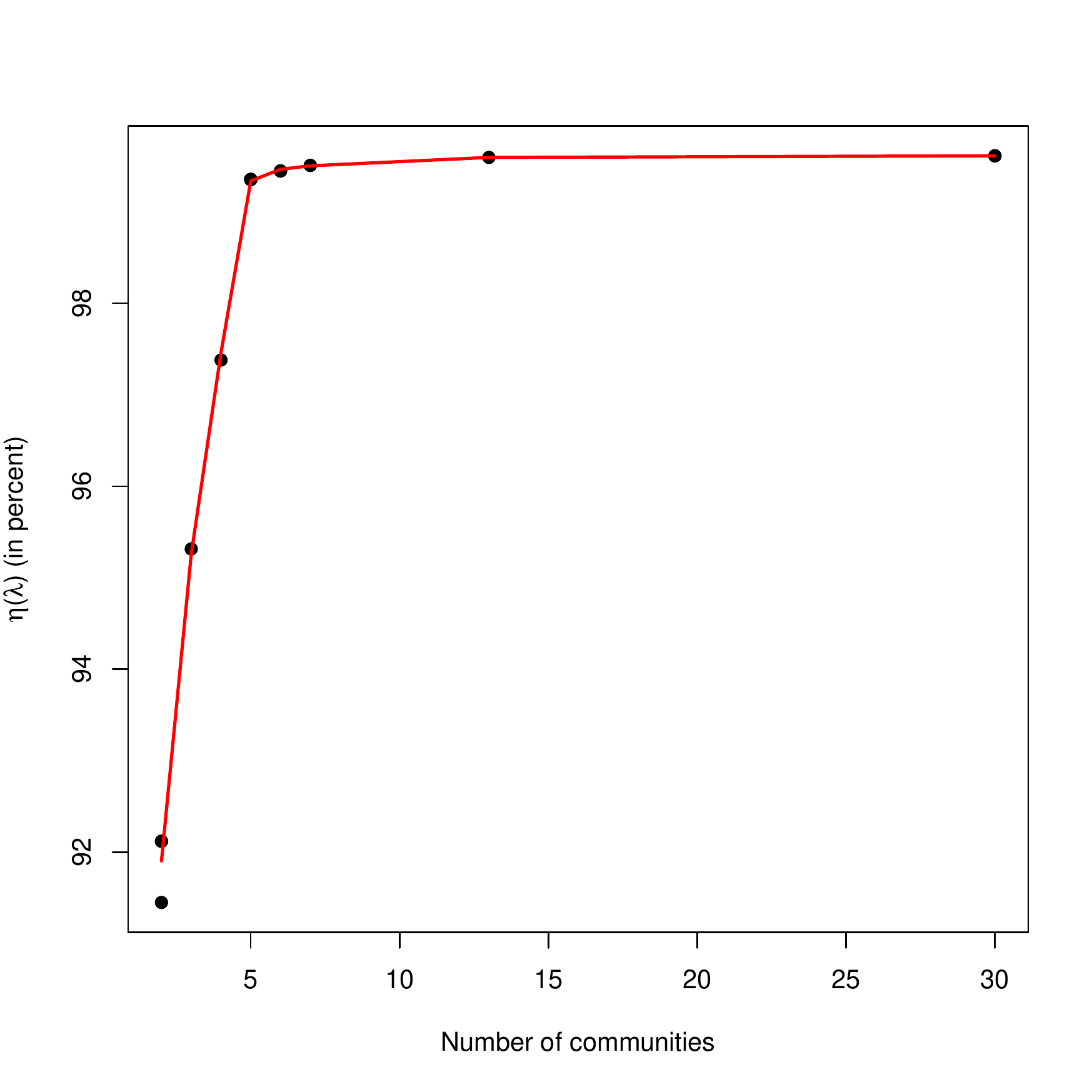}
        \end{minipage}
        \caption{Simulated data in which clear grouping exists.  $\eta(\lambda)$ vs number of communities plot flattens off at a prominent point. Curves with the same color are grouped together. The population can be accurately represented by a collection of group models.}
        \label{fig:clear_groups}
       
\end{figure*}

In lieu of analysis of operational fleet data, the results of the algorithm on synthetic data are illustrated in Fig. \ref{fig:no_groups}, \ref{fig:fuzzy_groups}, \ref{fig:clear_groups}.  
A population of 30 curves is analyzed. The independent variable of each curve (x) has been randomly sampled in the interval 0-100. The dependent variable of each curve (y) is generated using a linear model with additive gaussian noise having zero mean and variance of one. To state one application, the dependent variable may be the response of individuals to varying stimulus (independent variable) in a psychology experiment. 

The three cases considered are meant to verify whether we can use the algorithm to distinguish the cases where meaningful grouping exists from those where it does not and to find the groups in the latter case. Curves belonging to the same group have been assigned the same color. 

Figure \ref{fig:no_groups} represents the data generated by sampling the slope of each generative linear model from a uniform distribution in an appropriate interval. As a result, the curves are uniformly spread apart. The corresponding accuracy plot does not level-off as the number of communities increases. Hence, all the curves have been assigned different colors, implying that curves have not been grouped at all. 

Figure \ref{fig:fuzzy_groups} was generated from data where the slopes of the linear model were sampled from a mixture of 6 gaussian distributions with large variance. The plot suggests that the curves are spread apart, yet they fuzzily group together in pockets. The accuracy plot does not provide a single number for the clusters in the data. The color coding of the curves is shown for the case when the number of communities is 16. Curves in vicinity do share the same color, yet nothing conclusive can be said about the number of communities.

Figure \ref{fig:clear_groups} considers the case where the curves group themselves distinctly. The slopes were sampled from a mixture of 5 gaussian distributions with small variance. The accuracy plot has a distinct elbow for the number of community as 5. The corresponding grouping of the curves matches with the data.


\section{Discussion}

Grouping of entities in the aforementioned manner provokes new questions: why do units group themselves? What similarities do entities belonging to the same group exhibit? If meta-data about the entities is present, it may be possible to perform root-cause anlaysis and find the factors which cause such grouping. The detection of such factors may reveal previously unknown insights. These questions have been bypassed here by use of simulated data.

A suitable measure needs to be devised which can reflect the confidence of grouping by the algorithm. For example, our grouping has low confidence in Fig. 3, but high confidence in Fig. 4. Though such a conclusion can be drawn from the accuracy plot, a numerical value to suggest this will be convenient. Besides, the order/form of the generative curve models is typically unknown a priori. The model estimation procedure may result in overfitting or underfitting. The effect of model selection on the accuracy of the groups needs to be investigated.  

An alternative way of grouping regression models may be to cluster the coefficients of the models of individual units. For small datasets, the unreliability of coefficient estimates due to multicollinearity may be a challenge. Even if multicollinearity is absent, we need to factor the variable importance of coefficients, possibly by scaling the respective axes by variance values.

Grouping regression models may hold promise for other applications, such as reducing the number of trees in a random forest without compromising the radomness of the forest. Another application may be to analyse unbalanced datasets by sampling multiple datasets and analysing the relations among the resulting models. These applications will be explored in a later work.

\section*{Acknowledgment}
The authors would like to thank Subhankar Ghosh and Subodh Kolwankar for their valuable suggestions.

\bibliographystyle{plain}
\bibliography{grouping}

\end{document}